\documentclass[pra,superscriptaddress,twocolumn,floatfig,showpacs]{revtex4-1}
\usepackage{graphicx,color}
\usepackage{amsmath,amssymb,bm}
\usepackage{braket}		

\definecolor{darkred}{rgb}{0.90,0,0}
\definecolor{darkgreen}{rgb}{0,0.60,.2}
\definecolor{darkblue}{rgb}{0,0,1}
\definecolor{grey}{cmyk}{0,0,0,0.25}
\definecolor{orange}{cmyk}{0,0.6,1,0}

\begin{document}
\title{Temporal decay of N\'eel order in the one-dimensional Fermi-Hubbard model}

\author{A. Bauer}
\author{F. Dorfner}
\author{F. Heidrich-Meisner}
\affiliation{Department of Physics and Arnold Sommerfeld Center for Theoretical Physics,
Ludwig-Maximilians-Universit\"at M\"unchen, D-80333 M\"unchen, Germany}

\begin{abstract}
Motivated by recent experiments with ultra-cold quantum gases 
in optical lattices we study the decay of the staggered moment in the one-dimensional Fermi-Hubbard model
starting from a perfect N\'eel state using exact diagonalization and the iTEBD method. This extends previous work in which the same problem has been addressed
for pure spin Hamiltonians. As a main result, we show that the relaxation dynamics of the double occupancy 
and of the staggered moment are different. The former is controlled by the nearest-neighbor tunneling rate while the latter is much slower
and strongly dependent on the interaction strength, indicating that spin excitations are important. This difference in characteristic energy scales for the fast 
charge dynamics and the much slower spin dynamics is also reflected in the real-time evolution of nearest-neighbor density and spin correlations.
A very interesting time dependence emerges in the von Neumann entropy, which at short times increases linearly with a slope
proportional to the tunneling matrix element while the long-time growth of entanglement is controlled by spin excitations.
Our predictions for the different relaxation dynamics of the staggered moment and the double occupancy should be
observable in state-of-the art optical lattice experiments.
We further compare time averages of the double occupancy to both the expectation values in the canonical and diagonal ensemble,
which quantitatively disagree  with each other on finite systems.  We relate the question of thermalization
to the eigenstate thermalization hypothesis. 
\end{abstract}

\maketitle


\section{Introduction}
\label{sec:intro}
The non-equilibrium dynamics of order parameters in quenches from ordered into disordered phases
and vice versa has been the topic of many studies, including work on Bose-Einstein condensates \cite{miessner98,smith12},
bosons defined on lattice models \cite{braun15} and systems with antiferromagnetic order \cite{werner12,tsuji13}.
In quantum magnets, the dynamics of the staggered magnetization is a simple 
yet non-trivial example since the N\'eel state  is never an eigenstate of antiferromagnetic Heisenberg
models. 

In one spatial dimension, since the spontaneous breaking of a continuous symmetry is prohibited,
starting from a state with perfect N\'eel order, the staggered magnetization  is expected  to decay to zero under the unitary time evolution with a SU(2)-symmetric Hamiltonian. This problem
has been intensely studied for the spin-1/2 XXZ chain \cite{barmettler09,barmettler10,brockmann14,liu14,heyl14,fagotti14,wouters14,poszgay14,torres-herrera14}
and one observes a temporal power-law decay of the staggered magnetization to zero for the XX case
and indications of an exponential decay to zero in the interacting case \cite{barmettler09}.
The quantum quench dynamics starting from the N\'eel state has  attracted additional attention since an exact solution 
for the long-time asymptotic behavior could be obtained exploiting the integrability of the model
\cite{wouters14,poszgay14,liu14}. Therefore, the question of whether or not the steady state in this quench problem
can be described by the generalized Gibbs ensemble \cite{rigol07} could be addressed with rigor.

In the context of condensed-matter experiments, the decay of N\'eel order is related  to time-resolved spectroscopy with Mott insulators
in real materials \cite{wall11,ehrke11,rettig12}. 
In experiments with ultra-cold quantum gases, it is often particularly easy to prepare initial real-space product 
states with a high fidelity, which has been used as the starting point in several non-equilibrium studies
of Hubbard- and Heisenberg-type of models \cite{trotzky12,cheneau12,ronzheimer13,fukuhara13,fukuhara13a}.
The particular problem of the decay of N\'eel order has so far been addressed in the non-interacting
case in one dimension  \cite{pertot14} (where the initial state is an ideal charge density wave state of one spin component)  and for a two-dimensional system \cite{brown14}. Moreover, the 
decay of a spin spiral has been investigated in a two-component Bose gas in the strongly-interacting regime, where it can be  described by the Heisenberg model, in one and
two dimensions \cite{hild14}.
The reverse problem, namely the formation of 
antiferromagnetic order in time-dependent protocols is of equal relevance 
since this may provide a path for studying magnetic order in the quantum regime in ultra-cold atomic gas experiments \cite{demarco,deleo08,lubasch11,yi12,barthel09},
which has been the goal of a series of recent experiments \cite{joerdens08,schneider08,krauser12,greif13,hart14}.
For other  non-equilibrium experiments with fermions in optical lattices, see \cite{strohmaier,will14,schreiber15}.

In this work, we study the real-time decay of the N\'eel state  in the one-dimensional Fermi-Hubbard model, 
which, first, extends previous studies \cite{barmettler09} by incorporating charge dynamics 
and second, is motivated by  two related recent experiments with fermions in one dimension \cite{pertot14} and bosons in two dimensions \cite{brown14}.
The Hamiltonian reads
\begin{equation}
H = -t_0 \sum_{i} (c_{i+1,\sigma}^{\dagger}c_{i,\sigma} +h.c.) + U \sum_i n_{i\uparrow} n_{i\downarrow}  \label{hamiltonian} 
\end{equation} 
where $t_0$ is the hopping matrix element, $U$ is the onsite repulsion, $c_{i,\sigma}^{\dagger}$ creates
a fermion with spin $\sigma=\uparrow,\downarrow$ on site $i$ and $n_{i\sigma}= c_{i,\sigma}^{\dagger} c_{i,\sigma}$. 
The initial state is given by 
\begin{equation}
|\psi_0 \rangle = | \dots , \uparrow, \downarrow , \uparrow, \downarrow, \uparrow, \downarrow, \dots \rangle \,.
\end{equation}
Consequently, we are at half filling.
We use the  infinite-system size time-evolving-block-decimation (iTEBD) algorithm \cite{vidal07} to compute the time dependence of several observables such 
as the staggered magnetization, the double occupancy, nearest-neighbor correlations and the 
von Neumann entropy (we set $\hbar=1$). As a main result, we demonstrate that 
the relevant time scales for the relaxation of the double occupancy is set by the inverse of the hopping matrix element 
$1/t_0$ while for the staggered magnetization and nearest-neighbor spin correlations, the dynamics is the slower the larger $U$ is.
The difference in the relaxation dynamics can most clearly be discerned in the strongly-interacting regime  $U/t_0>4$.
This reflects the existence of  two characteristic velocities in the low-energy, equilibrium physics of strongly interacting one-dimensional systems, namely
the spin and charge velocity, related to 
spin-charge separation 
 \cite{giamarchi}.
Furthermore, there are fingerprints in the time dependence of the entanglement entropy. In general, in global
quantum quenches, one expects a linear increase of $S_{\rm vN}(t) \sim t$ in time \cite{dechiara06,laeuchli08,calabrese07}. 
In our case, we observe a short-time dynamics governed by charge excitations where $S_{vN} \sim t_0 t $ 
while at longer times $S_{vN} \propto   t/U $, suggesting that spin excitations are relevant for which the 
energy scale is the magnetic exchange constant $J=4 t_0^2/ U$.

Furthermore, we analyze the dependency of the double occupancy on the post-quench values of $U/t_0$ and
 we investigate whether  the steady-state values are  thermal or not. The latter is a possible scenario for an integrable 
1D model \cite{rigol07}. We observe that time averages are close to the expectation values in the diagonal ensemble \cite{rigol08}, 
while on the system sizes accessible to exact diagonalization, the expectation values in the diagonal and canonical ensemble are clearly different.
In this context, we also show that the distribution of eigenstate expectation values 
is in general broad, in contrast to systems that are expected to thermalize in the framework of the eigenstate thermalization hypothesis \cite{rigol08,srednicki94,deutsch91}.
The  observation of broad eigenstate expectation values of observables in our model is similar to those of 
Refs.~\cite{rigol09,santos10} made for integrable models of interacting spinless fermions.
For other recent studies of interaction quantum quenches in the one-dimensional Fermi-Hubbard model, see \cite{kollar08,ho10,genway12,hamerla13,iyer14,riegger15},
and for studies of  the time evolution starting from a perfect N\'eel state in higher dimensions, see \cite{werner12,tsuji13,queisser14}.
The non-equilibrium dynamics starting from this particular state yet combined with a sudden expansion
into a homogeneous empty lattice has been investigated in Ref.~\cite{vidmar13}.

The plan of this paper is  the following. We provide a brief overview over the numerical methods and definitions
in Sec.~\ref{sec:methods}. Section~\ref{sec:results} contains our main results,
discussing the time evolution of observables and von Neumann entropy,  steady-state values, 
thermalization, and the dynamics in the strongly interacting regime.
We conclude with a summary presented in Sec.~\ref{sec:conclude}.



\section{Numerical methods}
\label{sec:methods}

In this work we use two wavefunction-based methods, exact diagonalization (ED) and iTEBD, to study non-equilibrium dynamics in the Fermi-Hubbard model.
We further use a standard density matrix renormalization group code (DMRG) to compute ground-state expectation values \cite{white92,schollwoeck05}.

\subsection{iTEBD}
\label{sec:itebd}
We use Vidal's iTEBD algorithm for infinite systems to calculate the time evolution of the observables of interest starting from
the perfect N\'eel state. This method approximates the true wave-function by a matrix-product state ansatz \cite{schollwoeck11} appropriate for the 
thermodynamic limit and is related to time-dependent density matrix renormalization group methods \cite{daley04,white04} and TEBD for finite systems \cite{vidal04}. 
We use a  Trotter-Suzuki break-up of the time-evolution operator with a time step that is chosen small enough to resolve 
high-frequency oscillations at large $U/t_0$.
The maximum number of states is bounded by  $\chi_{\rm max}=1024$. We compared runs with different $\chi_{\rm max}$
and show only data for which the results are indistinguishable on the scale of the figures.

Compared to its siblings - the time-dependent density matrix renormalization group method \cite{daley04,white04} and TEBD \cite{vidal04} -
the advantage of iTEBD is clearly that it is set up directly for the thermodynamic limit. Moreover, both TEBD and iTEBD are particularly well-suited
for problems in which the initial state has an exact matrix-product state representation, which applies to our situation.
All theses approaches rely on approximating the time-evolved wave-function through  matrix-product states which only gives a faithful
representation if the time-evolved wave-function does not encode a large amount of entanglement \cite{schollwoeck11}. 
While our initial state is not entangled, the entanglement in a global quantum quench like ours grows linearly in time (see, e.g., \cite{dechiara06}),
which results in an exponential increase of computational effort \cite{schollwoeck11}. Thus, as the time evolution progresses, eventually,
going from time step $t$ to $t+\Delta t$ will consume more computational time than the whole previous calculation.
By keeping the discarded weight constant in every step, one accounts for the time-dependent increase of the entanglement entropy, and by carrying out simulations
with a different discarded weight the accuracy of the data can be controlled.
While the linear increase of the entanglement entropy with time is generic to a global quantum quench, the actual quench, model parameters and the observable determine
the actual numerical costs such that no general prediction of numerical effort and accuracy is possible.

\subsection{Exact diagonalization}
\label{sec:ed}

Our second method is exact diagonalization.
We perform the   time evolution in a truncated Krylov space (see \cite{manmana05} for a review and references).
To be able to treat larger systems we exploit symmetries of the Hamiltonian \eqref{hamiltonian}, namely conservation of total particle number $N$, total spin $S^z$, invariance under lattice translations (quasimomentum $k$), the parity and spin-flip symmetry.
In ED simulations, we use periodic boundary conditions, the number of sites is denoted by $L$.

\subsection{Observables}
Key quantities in our analysis are the double occupancy
\begin{equation}
d(t) = \frac{1}{L}\sum_{i=1}^{L}\langle n_{i\uparrow} n_{i\downarrow}\rangle,
\end{equation}
where the associated operator is $\hat d = \frac{1}{L}\sum_{i=1}^{L} n_{i\uparrow} n_{i\downarrow}$.
The staggered magnetization is
\begin{equation}
m_s(t) = \frac{1}{2L} \sum_{i=1}^L  (-1)^i \langle n_{i\uparrow} -n_{i\downarrow}\rangle\,.
\end{equation}
We further study nearest-neighbor density and spin correlations  defined as
$N_{i}= \langle n_i n_{i+1} \rangle$ and $S_{i} =\langle S^z_i S_{i+1}^z \rangle $, with 
$n_i = n_{i\uparrow}+n_{i\downarrow}$ and $S_i^z = (n_{i\uparrow}-n_{i\downarrow})/2$.
The von Neumann entropy for a central cut through the system
is computed from 
\begin{equation}
S_{\rm vN} = - \mbox{tr} \lbrack  \rho_A \ln \rho_A \rbrack,
\end{equation}
where $\rho_A$ is the reduced density matrix of one half of the system.


\section{Results}
\label{sec:results}

\begin{figure}[t]
\includegraphics[width=.96\columnwidth]{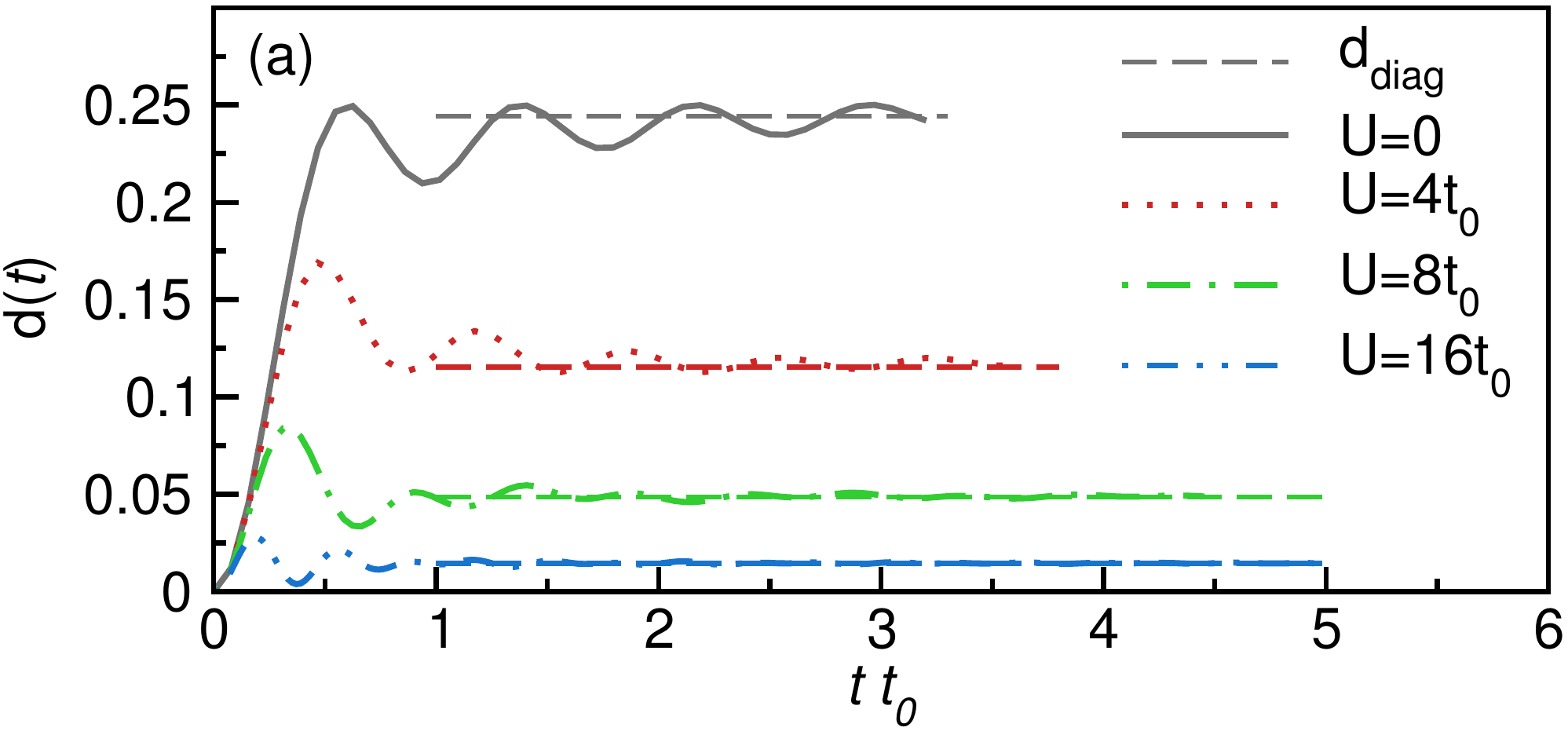}
\includegraphics[width=.96\columnwidth]{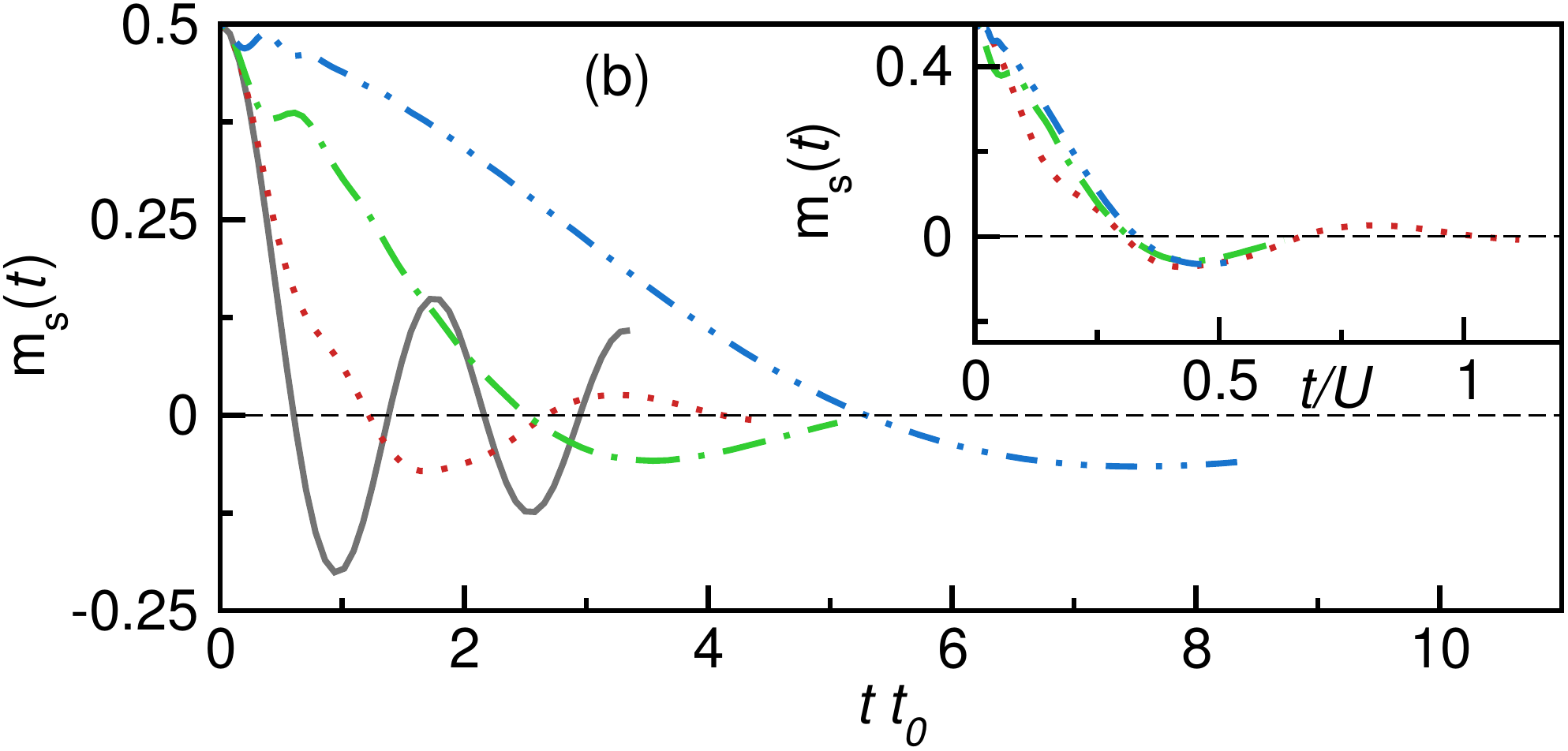}
\caption{(Color online) (a) Double occupancy $d(t)$ and (b) staggered magnetization  $m_s(t)$ as a function of time during the quench from the N\'eel state  to $U/t_0=0,4,8,16$ 
(iTEBD data).
Dashed lines in (a): expectation value $d_{\rm diag}$ in the diagonal ensemble Eq.~\eqref{eq:diag} from ED ($L=10$).
Inset in (b): $m_s(t)$ plotted versus $t/U$ for $U/t_0=4,8,16$.
}
\label{fig:mag}
\end{figure}

\subsection{Time evolution and characteristic time scales}
\subsubsection{Double occupancy and staggered moment}

Figures~\ref{fig:mag}(a) and (b) show the time evolution of the double occupancy $d(t)$ and of the staggered magnetization
$m_s(t)$, respectively, obtained from iTEBD simulations. While the double occupancy rapidly approaches a time-independent 
regime for all values of $U/t_0$ considered
here, the relaxation of the staggered magnetization towards $m_s=0$ is much slower. It is very instructive
to replot $m_s(t)$ versus $t/U$ [inset in Fig.~\ref{fig:mag}(b)]. This results in a collapse of the data for $U>4t_0$, which
is the better the larger $U/t_0$ is. Therefore, the  relaxation of double occupancy and staggered magnetization
occur at different time scales $1/t_0$ and $U$, respectively. This suggests that the  relaxation of spin-related quantities is 
set by the   magnetic exchange matrix element   given by $J=4t_0^2/U$ for large $U/t_0$.
Both quantities further exhibit coherent oscillations that decay during the approach to a stationary value.
For the double occupancy the frequency is given by $\omega=U$ for large $U\gg t_0$.
By contrast, the period of oscillations in $m_s(t)$ increases in the large $U/t_0$ limit. This is expected, since in the Heisenberg limit
the period of oscillations is  $1/(2J)$ with $J=4U/t_0^2$ \cite{barmettler09}.
Note that the non-interacting case has recently been studied comprehensively in \cite{sheikhan15} and that our iTEBD
results agree with the analytical solution for the $U=0$ case \cite{rigol06,barmettler09,cramer08}.

The short-time dynamics of both quantities ($\hat O$ representing an observable) can be obtained analytically by expanding the time-evolution operator:
\begin{eqnarray}
 \langle \hat O(t) \rangle &\approx & \langle \psi_0 | \hat O | \psi_0 \rangle + i \langle \psi_0| [H,\hat O]| \psi_0 \rangle t \nonumber \\ 
 && -\frac{1}{2}\langle \psi_0 |[H,[H,\hat O]]| \psi_0\rangle t^2+ \mathcal{O}(t^3)\,.
\label{eq:short}
\end{eqnarray}
For both double occupancy and staggered magnetization, the leading time dependence is $\sim t^2$ and comes from $\langle \psi_0  | H \hat O H | \psi_0 \rangle \propto t_0^2$, which is 
independent of $U$. Hence, the nontrivial $U$-dependence cannot be deduced from this short-time dynamics.
Second-order time-dependent perturbation theory in $t_0/U$ gives
\begin{eqnarray}
d(t) &=& \frac{8 t_0^2}{U^2} \sin^2\left(\frac{U t}{2}\right), \\
m_s(t) &=& \frac{1}{2} - \frac{8 t_0^2}{U^2} \sin^2\left(\frac{U t}{2}\right).
\end{eqnarray}
These expressions agree with our numerical data for $U/t_0 \gtrsim 16$.

\begin{figure}[t!]
\includegraphics[width=.96\columnwidth]{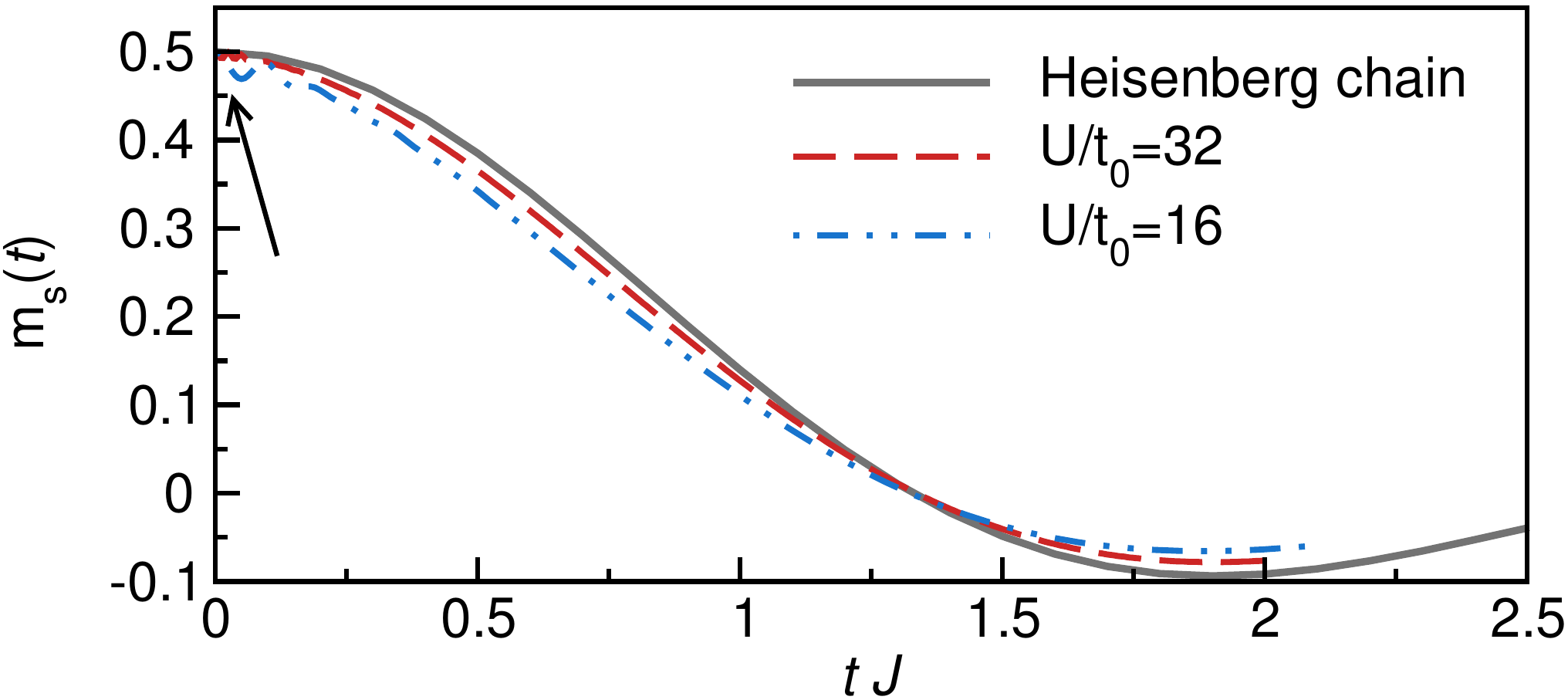}
\caption{(Color online)
Time dependence of the staggered moment for Heisenberg (solid line) and Hubbard model with a large $U/t_0=16,32$ (iTEBD data), plotted
versus time measured in inverse units of the magnetic exchange constant $J=4t_0^2/U$. The arrow indicates the small-amplitude oscillations in the short-time dynamics for finite $U/t_0<\infty$
whose frequency is given by $U$.}
\label{fig:heisenberg}
\end{figure}

\subsubsection{Comparison to Heisenberg model}
\label{sec:compheisenberg}
For completeness, we show that the time dependence of the staggered magnetization in the large $U/t_0$ limit  approaches 
 the one of the spin-1/2 Heisenberg model $H=J\sum_{i}\vec{S}_i \cdot \vec{S}_{i+1}$, where $J$ is the magnetic exchange coupling. 
We expect the time evolution of $m_s(t)$ to be identical in both models in the limit of large $U/J$ since the
Heisenberg model is derived from the Fermi-Hubbard model via a Schrieffer-Wolff transformation that projects onto the subspace of vanishing
double occupancy \cite{fazekas}.

The comparison is shown in Fig.~\ref{fig:heisenberg}, where we present iTEBD results for $U/t_0=16,32$
and the pure spin system. We see that the results for the two models become quantitatively similar for large $U/t_0$.  Moreover, the
short-time dynamics in $m_s(t)$, namely the small initial oscillations (see the arrow in the figure), disappear as $U/t_0$ increases, 
indicating the complete suppression of short-time charge dynamics. This is  accompanied by a shrinking of the time window in which  the short-time dynamics is 
governed by $\delta m_s(t)= m_s(t) -  m_s(t=0) \propto (t_0 t)^2$ [see Eq.~\eqref{eq:short}] which gets replaced by 
$ \delta m_s(t)  \propto (J t)^2$ (the latter follows from considering the Heisenberg model).

\subsubsection{Nearest-neighbor correlations}

\begin{figure}[t]
\includegraphics[width=.96\columnwidth]{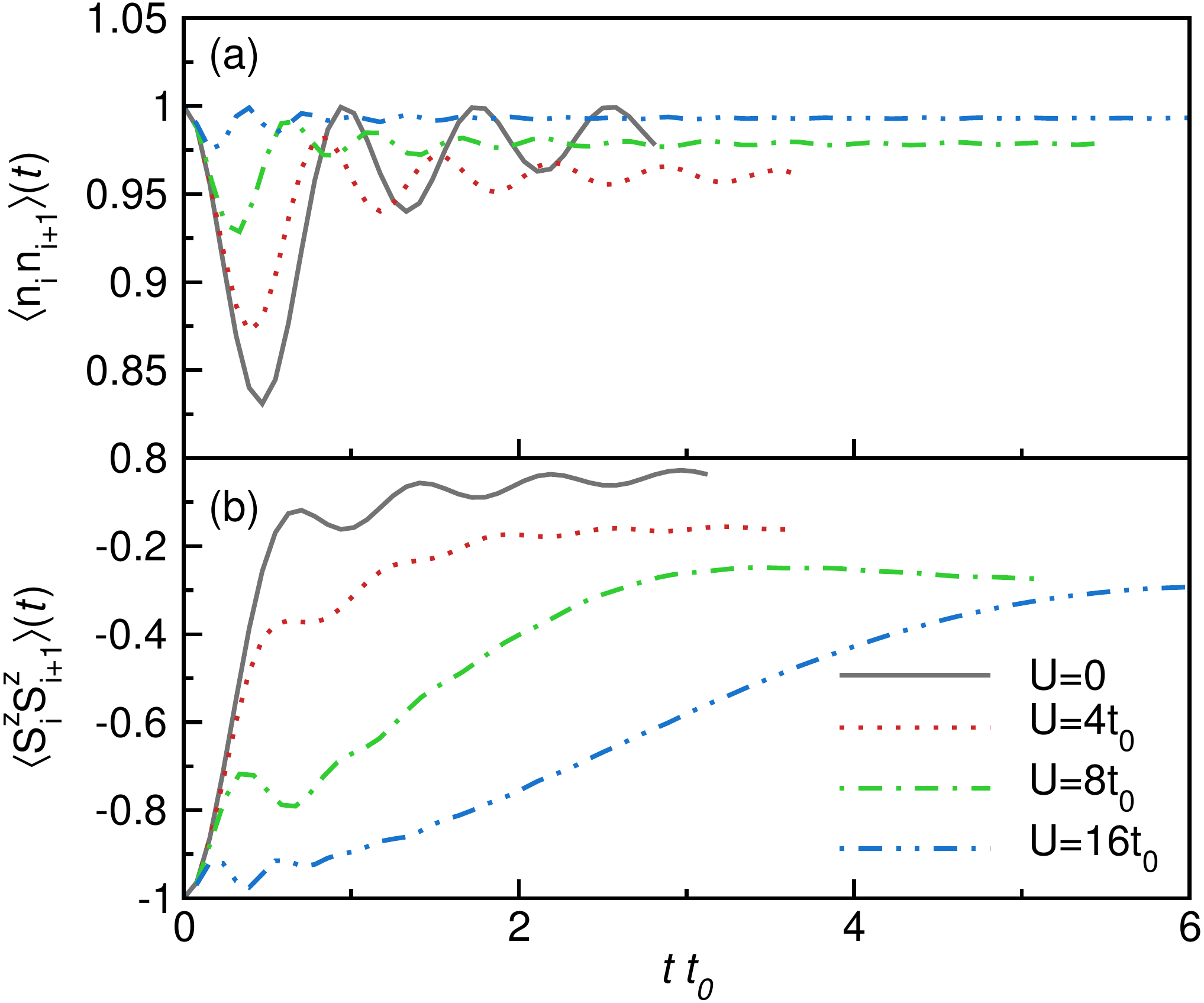}
\caption{(Color online) (a) Nearest-neighbor charge correlations $N_{i}(t) = \langle n_i n_{i+1}\rangle $ and (b) nearest neighbor, longitudinal spin correlations $S_i(t)=\langle S_i^zS_{i+1}^z\rangle $
as a function of time for  $U/t_0=0,4,8,16$ (iTEBD results).
}
\label{fig:corr}
\end{figure}

The time dependence of nearest-neighbor density correlations $N_{i}(t)$ [Fig.~\ref{fig:corr}(a)] and spin correlations $S_i(t)$ [Fig.~\ref{fig:corr}(b)],
bears similarities with  the one of the double occupancy and the staggered moment, respectively.
The density correlator undergoes 
a rapid decrease towards a stationary state that happens during the first tunneling time and then exhibits oscillations with a $U$-dependent frequency.
  On the contrary, the relaxation dynamics of the spin correlator is much slower, and again controlled by $U$ (the data for $S_i(t)$ 
can be collapsed in the $U/t_0>4$ regime by plotting them versus $t_0/U$, analogous to the staggered moment).

\subsubsection{Von Neumann entropy}

\begin{figure}[t]
\includegraphics[width=.96\columnwidth]{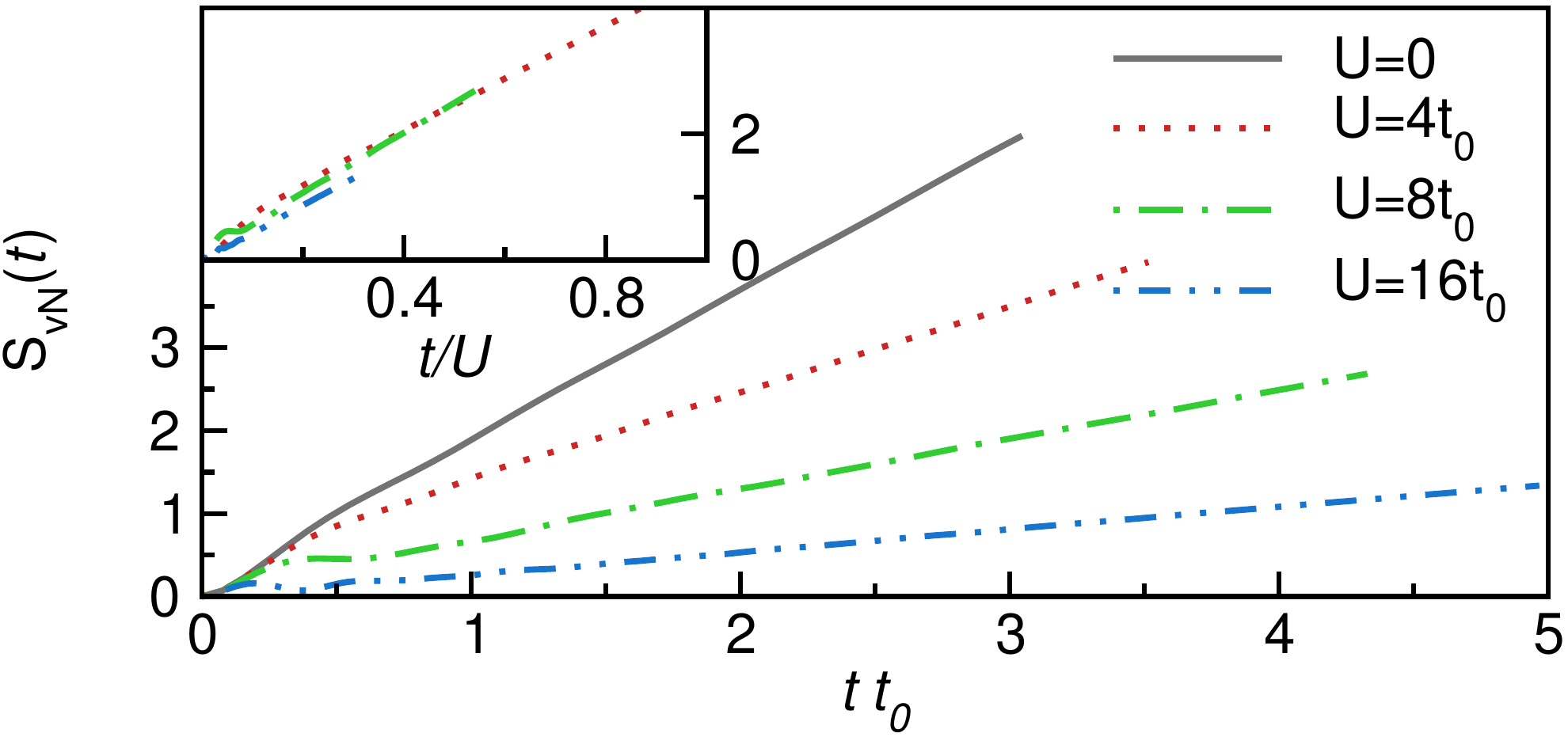}
\caption{(Color online) Von Neumann entropy $S_{\rm vN}$ for  a central cut through the system as a function of time for  $U/t_0=0,4,8,16$ (iTEBD results).
Inset: $S_{\rm vN}$ plotted versus $t/U$ for $U/t_0=4,8,16$.
}
\label{fig:svn}
\end{figure}

The existence of two different time scales for the relaxation dynamics of the double occupancy and the staggered magnetization
translates into an interesting time dependence of the von Neumann entropy (see Fig.~\ref{fig:svn}). At short times $t\lesssim 0.5/t_0$,
$S_{\rm vN} \sim t$ with a prefactor that is independent of $U$, while for $t\gtrsim 0.5 /t_0$, the time dependence crosses over to
a linear increase with a strongly $U$-dependent slope. Plotting $S_{\rm vN}$ versus $t/U$ results in a  collapse of the data [see the inset in Fig.~\ref{fig:svn}], 
comparable to the behavior of the staggered magnetization. 

\begin{figure}[b!]
\includegraphics[width=.96\columnwidth]{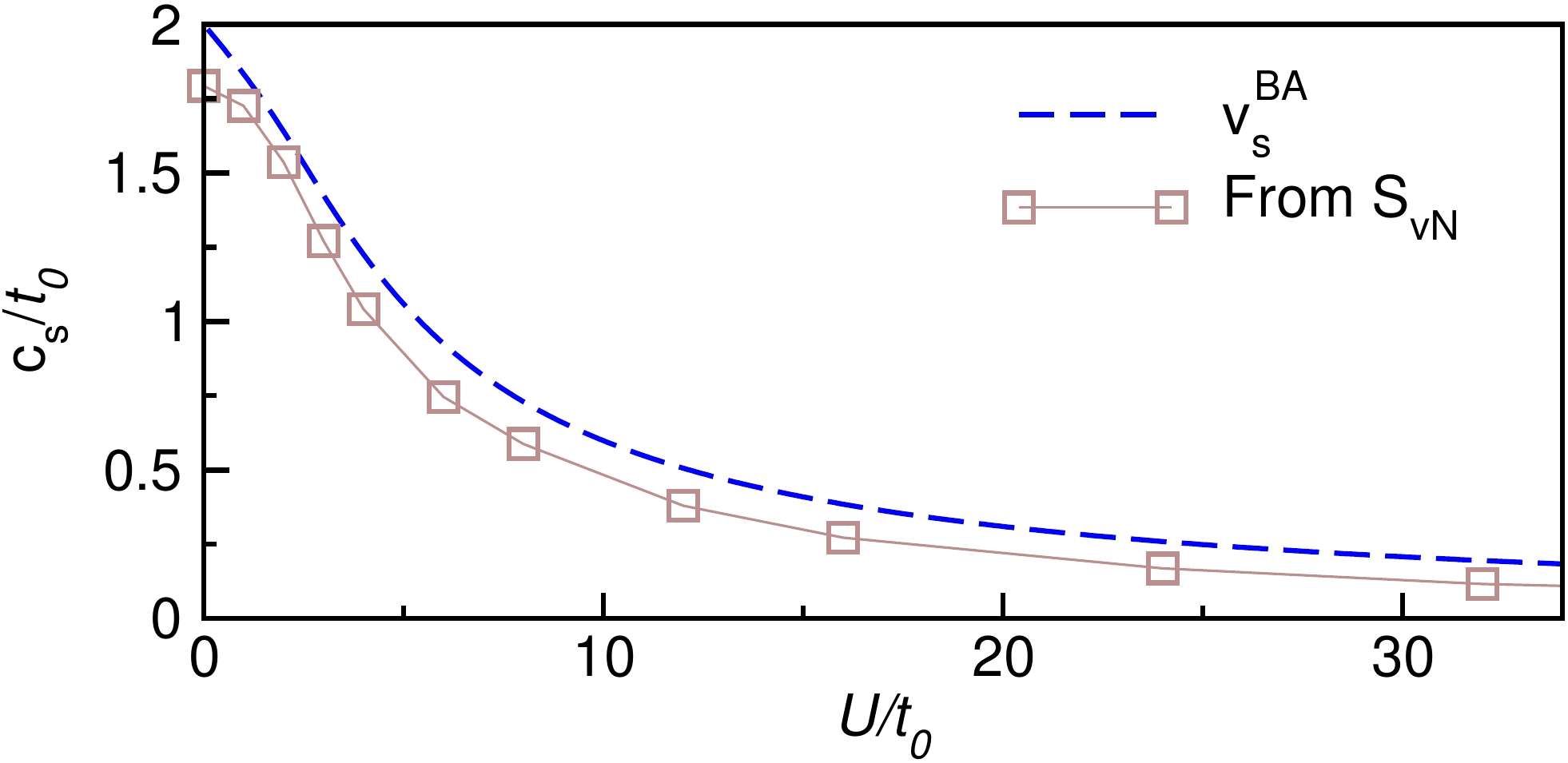}
\caption{(Color online) Characteristic velocities $c_s$ extracted from the time dependence of the von Neumann entropy $S_{\rm vN}$ 
in the $U$-dependent regime $t\gtrsim 0.5/t_0$, plotted versus $U/t_0$ (circles). For comparison we include the exact values $v_s^{\rm BA}$ (dashed line) of the 
spin velocity known from the Bethe-ansatz solution \cite{takahashi70}. 
} 
\label{fig:velocity}
\end{figure}

The prefactor $c_s$ of the linear increase of the von Neumann entropy is related to the existence of gapless modes and given by the 
characteristic velocities \cite{calabrese07}. We have extracted the prefactor of $S_{\rm vN}$ from the increase in the $U$-dependent regime, 
shown in Fig.~\ref{fig:velocity}. It turns out to be a monotonically decreasing function of $U/t_0$. We further
compare $c_s$ to the exact value of the spinon velocity $v_s^{\rm BA}$ known from the Bethe ansatz \cite{takahashi70,preuss94,essler-book} (dashed line in Fig.~\ref{fig:velocity}):
\begin{equation}
v_s^{\rm BA} = 2t_0 \frac{I_1(2\pi t_0/U)}{I_0(2\pi t_0/U)}\,
\end{equation}
($I_0$ and $I_1$ are modified Bessel functions of the first kind).
Both $c_s$ and $v_s^{\rm BA}$ clearly have a very similar  dependence on $U/t_0$, unambiguously showing that the long-time dynamics of the
entanglement entropy are controlled by  spin excitations.

\begin{figure}[t]
\includegraphics[width=.96\columnwidth]{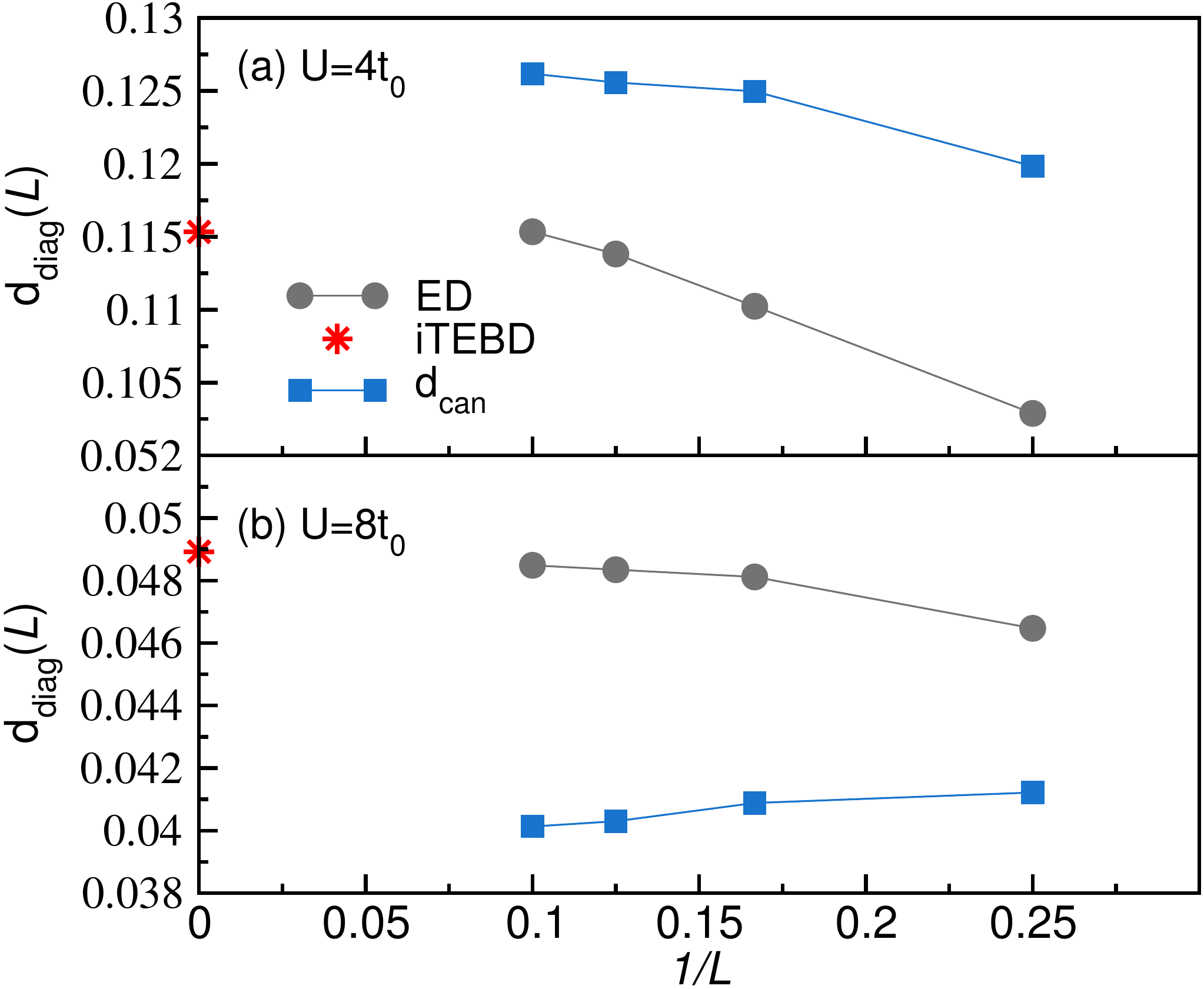}
\caption{(Color online)
Finite-size scaling of the expectation value of the double occupancy in the diagonal ensemble (circles) $d_{\rm diag}$
for (a) $U=4t_0$ and (b) $U=8t_0$ (circles: ED data for $L=4,6,8,10$, star: time average $\bar d$ from iTEBD). The figure also includes the
expectation values in the canonical ensemble $d_{\rm can}$ (squares).
}
\label{fig:d_L}
\end{figure}

\subsection{Time averages of double occupancy}
\label{sec:steadystate}

In the analysis of time averages, it is instructive to compare them to the expectation values in the diagonal and
canonical ensemble. The diagonal ensemble is defined as \cite{rigol08}
\begin{equation}
O_{\rm diag} = \sum_{\alpha} |c_{\alpha}|^2 \langle \alpha | \hat O | \alpha \rangle, 
\label{eq:diag}
\end{equation}
where $|\alpha\rangle$ are post-quench eigenstates ($H |\alpha\rangle = E_\alpha |\alpha\rangle$) and  $c_\alpha = \langle \psi_0 | \alpha \rangle$ are the overlaps
between the initial state and post-quench eigenstates. $O_{\rm diag}$ is the long-time average of $\langle \hat O \rangle $ \cite{rigol08},
where  degeneracies do not enter.

Given that the double occupancy can routinely be measured in quantum gas experiments \cite{strohmaier,ronzheimer13},
we concentrate the following discussion on this quantity.
The values for $d_{\rm diag}$ computed for $L=10$ using ED are included in Fig.~\ref{fig:mag}(a) as dashed lines. 
Clearly, the time-dependent iTEBD data
are very close to  $d_{\rm diag}$ and seem to approach this value as the amplitude of oscillations decays.

To get a feeling for the system-size dependence,   we show $d_{\rm diag}$  versus $1/L$ for (a) $U=4t_0$  and (b) $U=8t_0$, together with $\bar d$
extracted from iTEBD simulations plotted at $1/L=0$ in Figs.~\ref{fig:d_L}(a) and (b), respectively. 
The finite-size dependence of the data for $d_{\rm diag}$ is consistent with $d_{\rm diag}(L) \to \bar d$ as system size increases. 
  We should stress, though, that the time average of the double occupancy itself 
 could change if we were able to reach longer times with iTEBD.

\begin{figure}[t]
\includegraphics[width=.96\columnwidth]{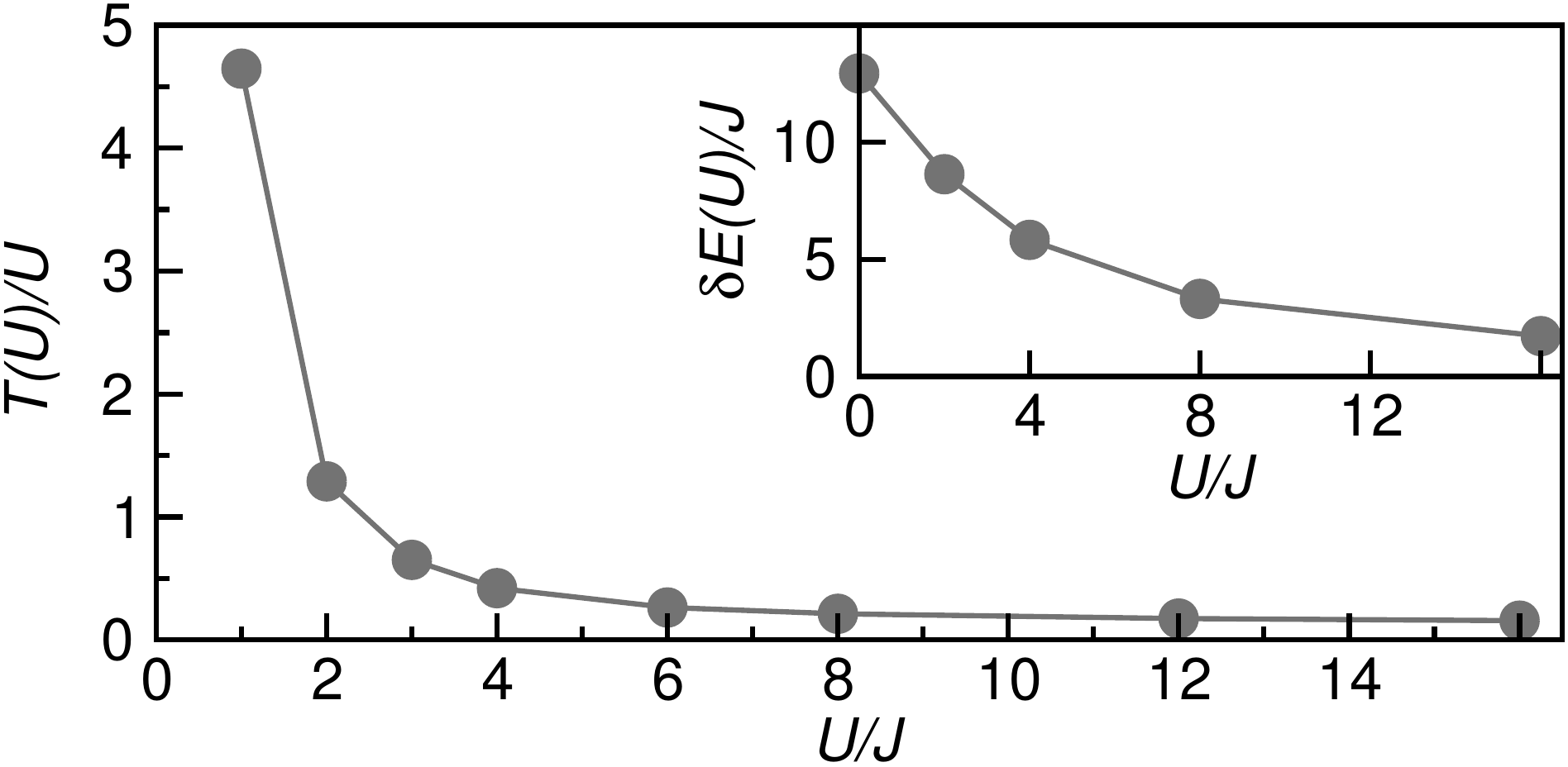}
\caption{(Color online)
Canonical temperature $T/U$ (main panel) and excess energy $\delta E$ (see Eqs.~\eqref{eq:canT} and \eqref{eq:deltaE})
versus $U/J$ for $L=10$ (ED results).
}
\label{fig:T}
\end{figure}

\begin{figure}[t!]
\includegraphics[width=.96\columnwidth]{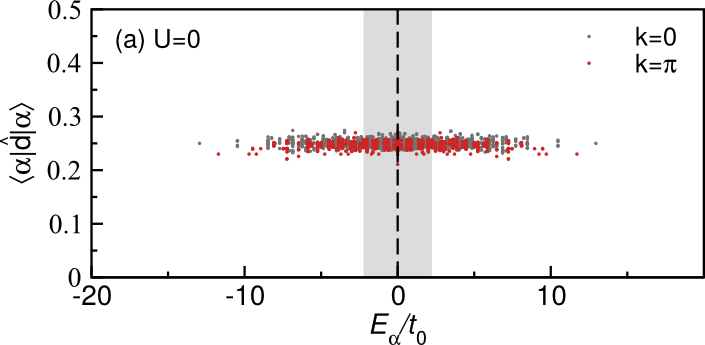}
\includegraphics[width=.96\columnwidth]{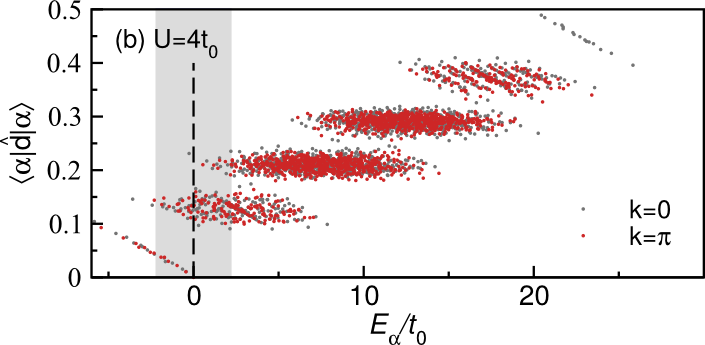}
\includegraphics[width=.96\columnwidth]{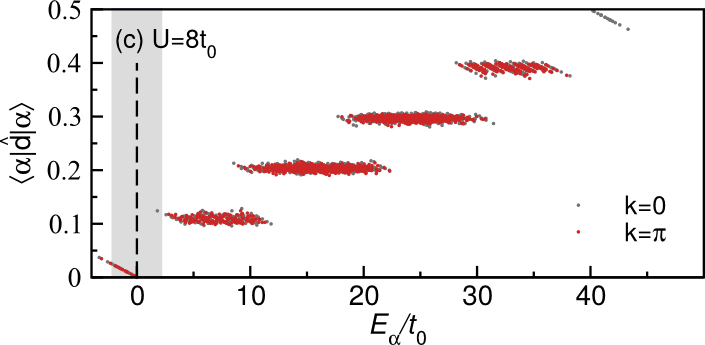}
\includegraphics[width=.96\columnwidth]{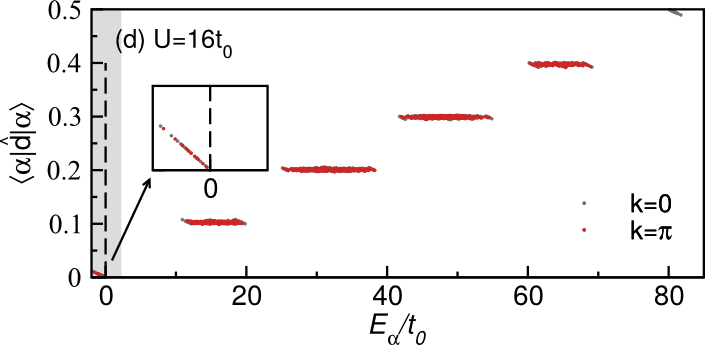}
\caption{(Color online) Post-quench eigenstate expectation values for the double occupancy for various interaction strengths: (a) $U/t_0=0$,  (b) $U/t_0=4$, (c) $U/t_0=8$, (d) $U/t_0=16$ (ED data for $L=10$).
The vertical dashed lines mark the quench energy $E=0$ for our initial state.
The inset in (d) shows a blow-up of the first doublon band.
The N\'eel state is doubly degenerate (denoted by $|\psi_0\rangle $ and $|\tilde\psi_0\rangle $)
and the linear combinations $|\psi_{\pm}\rangle = (|\psi_0\rangle  \pm|\tilde\psi_0\rangle)/\sqrt{2}$
live in the total quasi-momentum $k=0$ and $k=\pi$ subspaces. 
}
\label{fig:ethcloud}
\end{figure}

The expectation value in the canonical ensemble is computed from 
\begin{equation}
O_{\rm can} = \mbox{tr} [\rho \hat O]\,, \label{eq:can}
\end{equation}
where $\rho=\mbox{exp}(-\beta H)/Z$ with $Z$ the partition function, all evaluated at fixed $N=L$ and vanishing total spin $\sum_{i=1}^L \langle S_i^z \rangle=0$. The temperature $T=1/\beta$ is fixed by
requiring that 
\begin{equation}
E=\langle \psi_0 | H |\psi_0 \rangle = \mbox{tr} [\rho H] \,.
\label{eq:canT}
\end{equation}
While in  our problem $E=0$, independently of the post-quench value of $U$, the canonical temperature 
$T$ clearly is a function of $U/J$ since the post-quench ground-state energy $ E_{\rm gs}(U)$, defining for each $U/J$ the 
zero-temperature reference point, depends on $U/J$. To illustrate this point, we introduce the   
  {\it excess}  energy 
\begin{equation}
\label{eq:deltaE}
\delta E=E - E_{\rm gs}(U)\,.
\end{equation}
The canonical temperature $T/U$ expressed in units of $U$ and the excess energy $\delta E$ are plotted versus $U/J$ in the main panel and inset of Fig.~\ref{fig:T}, respectively.
Both $T/U$ and $\delta E$ are monotonously increasing functions of $U/J$ as $U/J$ is lowered. At $U/J=\infty$, $\delta E$ is zero since the initial state is in the ground
state manifold in that limit. As $U/J$ decreases, $E=0$ moves towards the middle of the many-body spectrum (see also the discussion in Sec.~\ref{sec:props} and Fig.~\ref{fig:ethcloud})
and eventually, at $U=0$, it translates into an infinite temperature  (see also \cite{sorg14}).
It is thus more appropriate to express $T$ in units of $U$ rather than $J$ in the large $U/J$ regime since this results in $T/U \to 0$ for $U/J\to \infty$ \cite{sorg14}. 

\begin{figure}[t]
\includegraphics[width=.96\columnwidth]{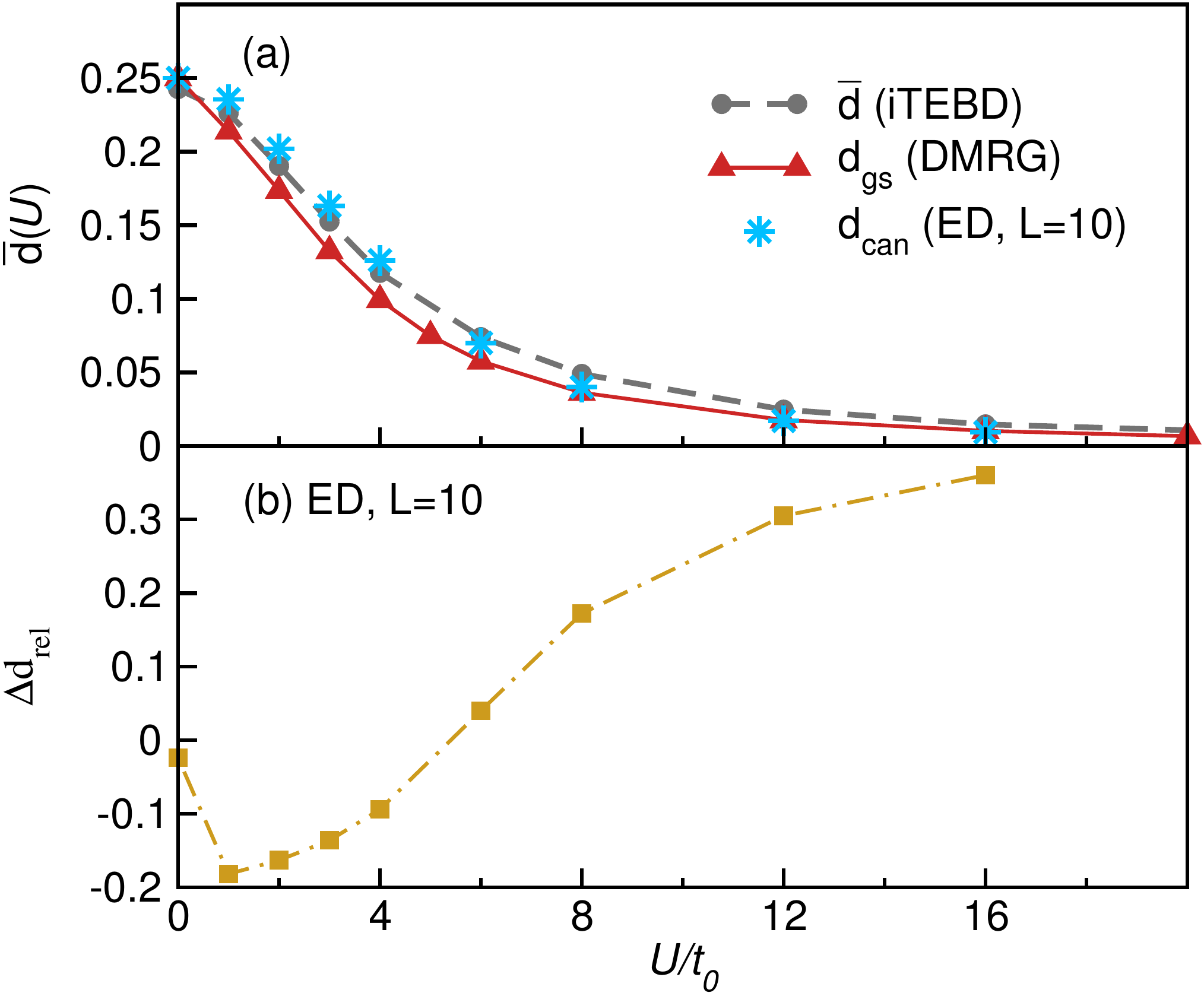}
\caption{(Color online) (a) Time averages (circles) of  the double occupancy  as a function of $U/t_0$.
Time averages are obtained by averaging over full periods of the oscillations.
 Triangles denote the
ground-state expectation values (computed with DMRG for $L=64$ and open boundary conditions) for comparison, stars are the expectation values in the
canonical ensemble Eq.~\eqref{eq:can} computed with ED for $L=10$.
(b) Relative difference between canonical and diagonal ensemble $\Delta d_{\rm rel}= (d_{\rm diag} -d_{\rm can})/d_{\rm diag}$.}
\label{fig:davg}
\end{figure}

From the time-dependent data shown in Fig.~\ref{fig:mag}(a), we extract the time averages $\bar d$ of the double occupancy.
These are displayed in Fig.~\ref{fig:davg}(a) versus $U/t_0$ (circles) together with the expectation values $d_{\rm gs}$
in the ground state (triangles, DMRG data) and the expectation value $d_{\rm can}$ in the canonical ensemble (stars).
First, we observe that, as anticipated from Fig.~\ref{fig:mag}(a), $\bar d \approx d_{\rm diag}$ for the accessible time scales/system sizes [data for $d_{\rm diag}$
not shown in  Fig.~\ref{fig:davg}(a)].

Second, the time averages are {\it above} the ground-state expectation values. 
This behavior is, in the large $U/t_0$ limit, somewhat unexpected, given the known  non-monotonic temperature dependence
of $d$.   
As a function of $T$, the equilibrium double occupancy $d(T)$ first decreases from its zero-temperature value and then increases for large $T$ towards $d(T=\infty)=1/4$ (see \cite{gorelik10,gorelik12}).
The position of the minimum in $d(T)$ can be interpreted as a scale for the  separation of the spin- versus charge excitation dominated temperature regime. 
Since we do not observe $\bar d < d_{\rm gs}$ up to $U/t_0=64$, we conclude that the initial state always mixes in 
doublons from the upper Hubbard band and not just the virtual doublons present in the ground state.
For the accessible system sizes, this is confirmed by the discussion presented in Sec.~\ref{sec:ensemble}.

We further observe the known $d_{\rm gs}\propto 1/U^2$ behavior \cite{shiba66,pelizzola} in the large $U/t_0$ regime  (also obeyed by $\bar d$).
The  value of $d=1/4$, which is the 
infinite-temperature expectation value at $U=0$,  is approached by $\bar d$ and $d_{\rm diag}$ as $U/t_0$ is lowered (see Fig.~\ref{fig:davg}).

Since the system is integrable, it is not surprising that 
the expectation values in the canonical ensemble are different from the ones in the diagonal ensemble. 
The canonical ensemble has been computed for a small system using exact diagonalization, and therefore, 
a quantitative comparison only makes sense by comparing to the diagonal ensemble but not to the iTEBD time averages. The relative difference is shown in Fig.~\ref{fig:davg}(b) for $L=10$
and can be quite large. 
At least for the accessible system sizes (see Fig.~\ref{fig:d_L}), this difference does not seem to become smaller.
Therefore, we do not observe thermalization in this model for the quench protocol studied here.
Nonetheless, the qualitative dependence of $\bar d$, $d_{\rm can}$ and $d_{\rm diag}$ on $U/t_0$ is quite similar.

\subsection{Connection to eigenstate thermalization hypothesis}
\label{sec:ensemble}

\subsubsection{Eigenstate expectation values}

One popular framework to understand   thermalization in closed many-body systems is the eigenstate thermalization hypothesis (ETH) \cite{rigol08,deutsch91,srednicki94}.
It states that $O_{\rm diag} = O_{\rm mc}$, where $ O_{\rm mc}$ is the expectation value in the micro-canonical ensemble, if 
the expectation values $\langle \alpha | \hat O | \alpha\rangle $ of $\hat O$ (a local observable) in post-quench eigenstates  
only depend on energy $E$ in the thermodynamic limit (the latter also assuming a narrow initial state \cite{rigol08,sorg14}).
In other words, expectation values computed in a typical many-body eigenstate (which should be the vast majority of all states)
already yield thermal behavior. For sufficiently large systems, expectation values in the microcanonical and canonical ensemble should agree with each other.

On a finite system accessible to exact diagonalization, validity of the ETH  manifests itself in a narrow width of  $\langle \alpha | \hat O | \alpha\rangle$
at a fixed energy $E$ for a generic quantum system, while for a 1D integrable system, $\langle \alpha | \hat O | \alpha\rangle$ can be very broad
for a given energy, due to the existence of many non-trivial (local) conservation laws resulting in a large fraction of degeneracies.
The picture has been studied and often verified  (see, e.g., \cite{rigol08,rigol09a,rigol09,rigol10,santos10,roux10,genway12,steinigeweg13,kim14}), the important question being how quickly 
the distributions of $\langle \alpha | \hat O | \alpha\rangle$ become sufficiently narrow as system size increases. Recent work suggests that for 
a generic system, this is exponentially fast in $L$ (\cite{beugeling14}, see also \cite{steinigeweg14,sorg14}), while for an integrable system,
the decay of the width of $\langle \alpha | \hat O | \alpha\rangle$ at a given $E$ is at most power-law \cite{ikeda13,beugeling14,alba15} (see also \cite{biroli10}).

Here, we exclusively analyze the distribution of post-quench eigenstate-expectation values of the double occupancy. These are presented in 
Figs.~\ref{fig:ethcloud}(a)-(d) for $U/t_0=0,4,8,16$.
For $U/t_0 \gtrsim 4$, the distributions have a very regular structure inherited from the $U/t_0=\infty$ limit, where the double occupancy is 
a conserved quantity. There is one band  for each possible value of $\langle \alpha |\hat d| \alpha \rangle$ (for the parameters of the
figure, $L=10$ these are $L \langle \alpha |\hat d| \alpha \rangle =0,1,2,3,4,5$).
For a nonzero and small $t_0/U$, the exact degeneracy in these bands is lifted while the structure as such is preserved on these
small systems. In the lowest  band, the effect of $t_0\not=0$ is to lower the energy from the
degenerate $U/t_0=\infty$ ground-state manifold at $E=0$ towards the correlated ground state, resulting at the same
time in an {\it increase} of $\langle \alpha |\hat d| \alpha \rangle$ towards its nonzero ground-state expectation value. 
This lowest band is very sharp and its negative slope translates into the  decrease of $d=d(T)$ from its zero-temperature value as a function
of temperature at low $T$ \cite{gorelik10,gorelik12}, which persists as long as the $dL =0$ band remains well separated from the $dL=1$ band.

At smaller $U/t_0$, the bands eventually start to overlap and they become very broad at a fixed energy (compare the discussion in \cite{rigol10,santos10a} for other models). At $U=0$, 
the distribution of $\langle \alpha |\hat d| \alpha \rangle$ becomes flat, resulting in an essentially energy-independent  mean value of 
$ \overline{\langle \alpha |\hat d| \alpha \rangle}\approx 1/4$.

\subsubsection{Properties of the specific initial state}
\label{sec:props}
Our initial state has a mean energy of $E=0$ and a width (in the diagonal ensemble) 
of $\sigma_{\rm diag} = \sqrt{\langle \psi_0 | H^2 | \psi_0 \rangle } =  t_0\sqrt{2L}$, which is independent
of $U$. This is indicated by the shaded areas in Fig.~\ref{fig:ethcloud}.
For large $U/t_0$, primarily the very narrow first band is sampled and $E=0$ sits at the high-energy edge of the first, $dL=0$ band (recall that for $U/t_0=\infty$, $dL$ takes integer
values).
Therefore, the initial state asymmetrically mixes in eigenstates with too large values of $\langle \alpha |\hat d| \alpha \rangle$ from first, states in the $dL=0$
band at $E<0$ and second, from the band with $dL=1$ (the latter follows from analyzing the distribution of $|c_{\alpha}|^2$).
Hence, the overall structure of the distribution of $\langle \alpha |\hat d| \alpha \rangle$ combined with the distribution of $|c_{\alpha}|^2$
is  consistent with the
observation that $\bar d> d_{\rm can}$ at large $U/t_0$ (compare Sec.~\ref{sec:steadystate}).
 
At very small $U/t_0$, the initial state samples the bulk of the system where the density of states is large. At $U=0$, the corresponding
canonical temperature derived from the quench energy is infinite and since $\langle \alpha |\hat d|\alpha \rangle$ does not depend much on energy, we must find $\bar d =d_{\rm diag}= d_{\rm can}\to 1/4$ as $L$ increases, 
consistent with the discussion in  Sec.~\ref{sec:steadystate}.
At intermediate $U/t_0$, the initial state samples several overlapping and partially very broad bands of the $\langle \alpha |\hat d| \alpha \rangle$ distribution [see, e.g., the case of $U/t_0=4$ shown in Fig.~\ref{fig:ethcloud}(b)]. Therefore, based on the structure of the eigenstate-expectation-value distributions  at the quench energy, we
 expect deviations between thermal behavior at intermediate and large $U/J$, consistent with our previous analysis. In conclusion, we stress that the quench energy alone is not
a sufficient criterion for the analysis of finite-system size data, but that the actual distribution of overlaps $|c_{\alpha}|^2$ crucially determines which bands are involved (see also the discussion in \cite{sorg14}).

\section{Summary and Conclusion}
\label{sec:conclude}
In this work, we studied the relaxation dynamics in the one-dimensional Fermi-Hubbard model starting from a perfect N\'eel state 
as a function of the interaction strength $U/t_0$. As a main result, we reported evidence that the 
relaxation dynamics of the staggered moment, spin correlations and of the von Neumann entropy at long times 
is controlled by spin excitations, while the double occupancy undergoes a much faster dynamics controlled by charge excitations.
The slope $c_s$ of the increase of the von Neumann entropy $S_{\rm vN} = c_s t$ is very similar to the
exact spinon velocity known from the Bethe ansatz.
This separation of time scales for double occupancy versus staggered magnetization could be accessible in 
state-of-the-art quantum gas experiments.  

We further demonstrated that the time averages of the double occupancy
are different from the expectation values in the canonical ensemble.
Nonetheless, both quantities exhibit the same  qualitative
dependence on $U/t_0$. 
Finally, we made a connection to the eigenstate thermalization hypothesis by showing that the eigenstate expectation values
of the double occupancy are, in general, broadly distributed with no well-defined dependence on energy only, characteristic for an integrable 
one-dimensional system.

{\it Acknowledgment.}
We thank F. Essler, M. Rigol, U. Schneider, D. Schuricht, and L. Vidmar  for helpful discussions.
F.D. and F.H.-M. acknowledge support from the Deutsche Forschungsgemeinschaft (DFG) via 
Research Unit FOR 1807 through grant no. HE~5242/3-1.

\bibliography{references}

\end{document}